\documentstyle[aps,multicol,epsf,citesort]{revtex}
\draft
\newcommand{\beq}{\begin{equation}}
\newcommand{\eeq}{\end{equation}}
\newcommand{\bdis}{\begin{displaymath}}
\newcommand{\edis}{\end{displaymath}}
\newcommand{\bea}{\begin{eqnarray}}
\newcommand{\eea}{\end{eqnarray}}
\newcommand{\barr}{\begin{array}}
\newcommand{\earr}{\end{array}}

\begin{document}
\title
{Liquid State Anomalies for the \\Stell-Hemmer Core-Softened Potential}

\author{ M. Reza Sadr-Lahijany, Antonio Scala, Sergey~V.~Buldyrev,
and H. Eugene Stanley}

\address{Center for Polymer Studies and Department of Physics,
        Boston University,\\Boston, Massachusetts 02215.}
\date{May 15, 1998}

\maketitle

\begin{abstract}

We study the Stell-Hemmer potential using both analytic (exact $1d$ and
approximate $2d$) solutions and numerical $2d$ simulations. We observe
in the liquid phase an anomalous decrease in specific volume and
isothermal compressibility upon heating, and an anomalous increase in
the diffusion coefficient with pressure. We relate the anomalies to the
existence of two different local structures in the liquid phase. Our
results are consistent with the possibility of a low temperature/high
pressure liquid-liquid phase transition.

\end{abstract}
\date{ssbs.tex ~~~ May 15, 1998 ~~~ draft}

\begin{multicols}{2}

In their pioneering work, Stell and Hemmer proposed the possibility of a
new critical point in addition to the normal liquid-gas critical point
for potentials that have a region of negative curvature in their
repulsive core (henceforth referred to as core-softened
potentials)~\cite{Stell}. They also pointed out that for the $1d$ model
with a long long range attractive tail, the isobaric thermal expansion
coefficient, $\alpha_P\equiv V^{-1}\left(\partial V/\partial T\right)_P$
(where $V,T$ and $P$ are the volume, temperature and pressure) can take
an anomalous negative value. Debenedetti et al., using thermodynamic
arguments, pointed out that the existence of a ``softened core'' can
lead to $\alpha_P<0$~\cite{Debenedetti}.

Here we further investigate properties of core-softened potential
fluids. We first study the properties of the $1D$ fluid using an exact
solution. We then investigate the behavior of the $2D$ fluid, initially
by an approximate solution provided by cell theory method and finally
by performing molecular dynamics simulation of the fluid.

\begin{figure}[htb]
\narrowtext
\centerline{
\hbox  {
        \epsfxsize=4cm
        \epsfbox{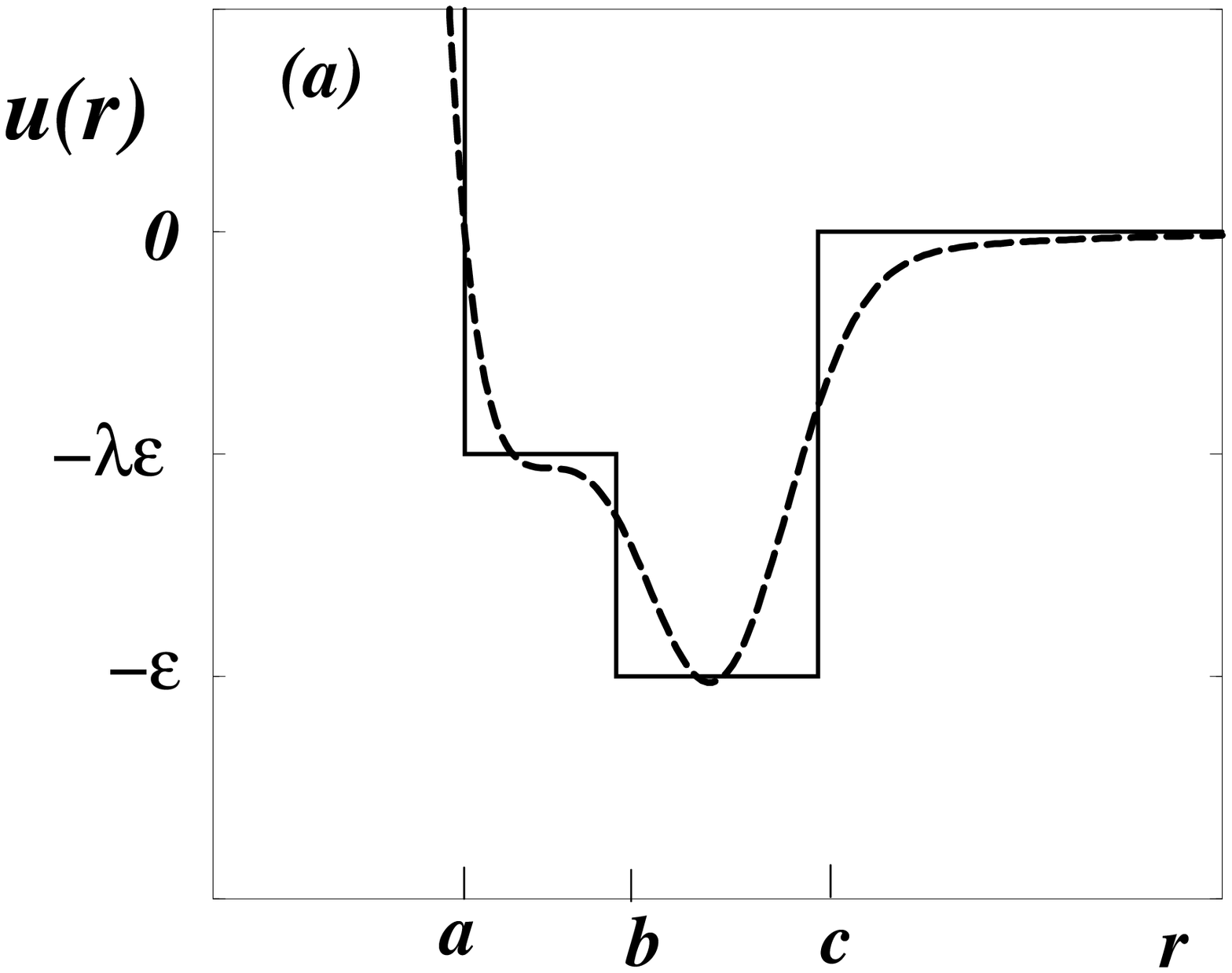}
        \hspace*{0.1cm}
        \epsfxsize=4cm
        \epsfbox{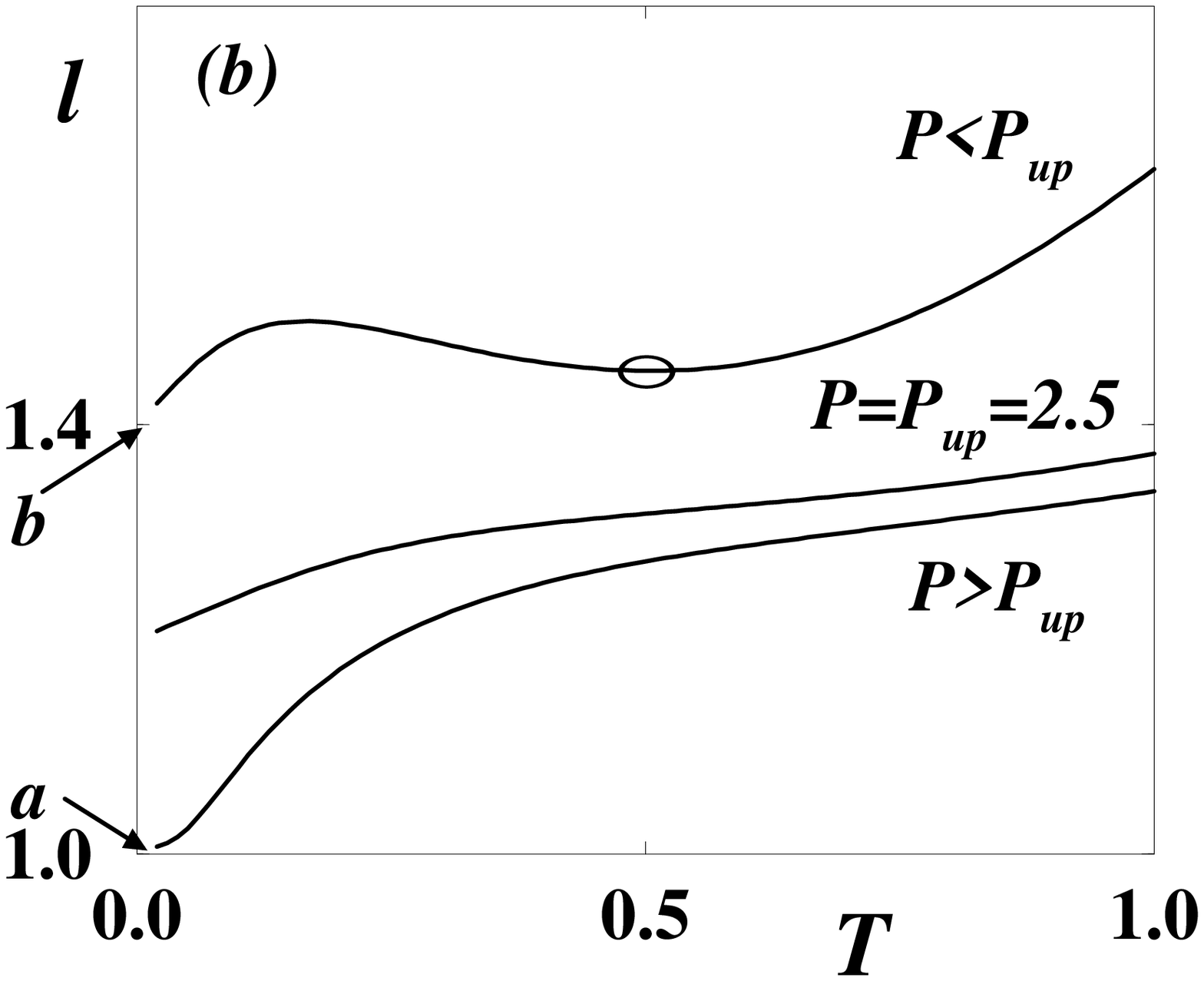}
       }
          }    

\centerline{
\hbox  {
        \epsfxsize=4cm
        \epsfbox{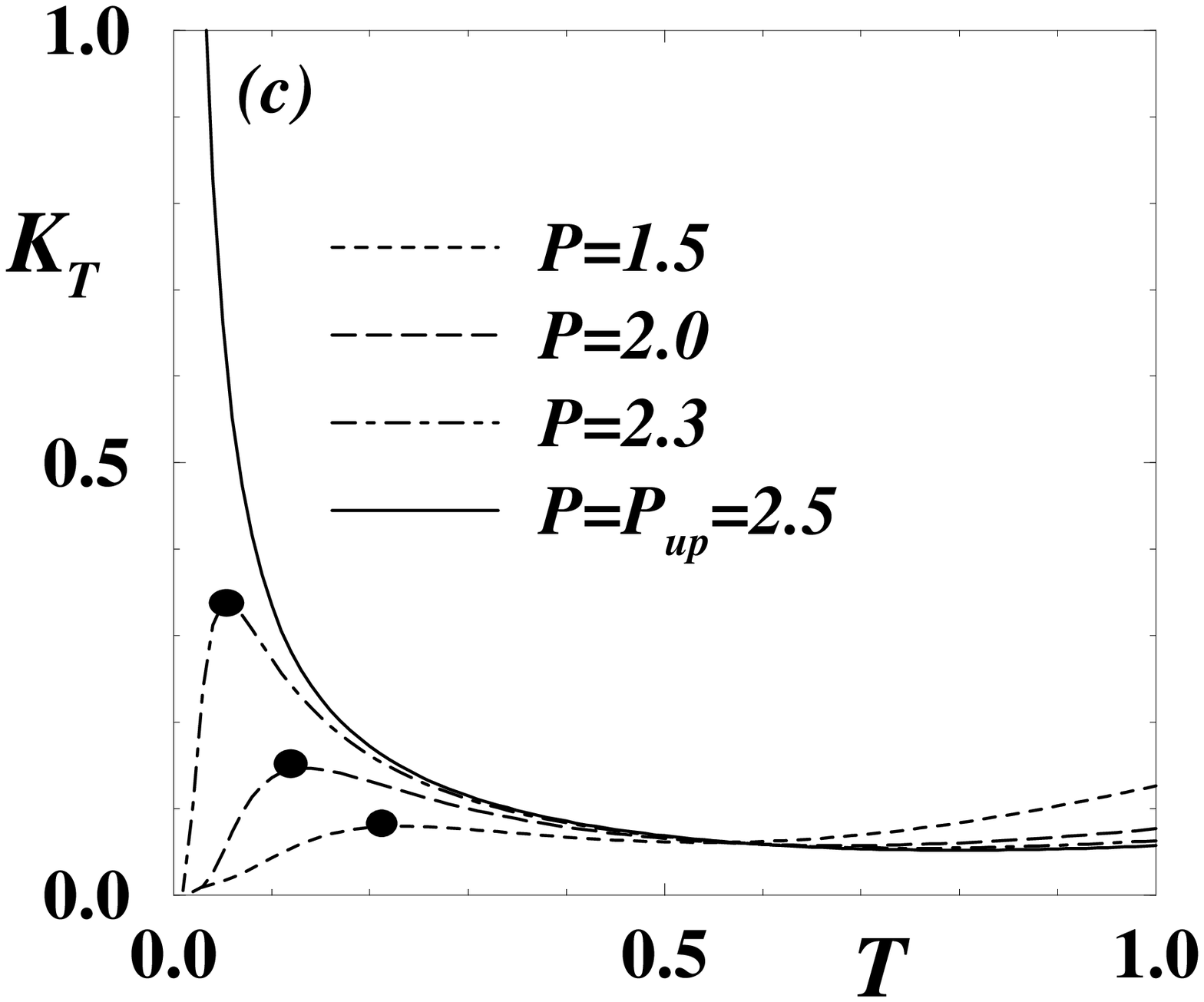}
        \hspace*{0.1cm}
        \epsfxsize=4cm
        \epsfbox{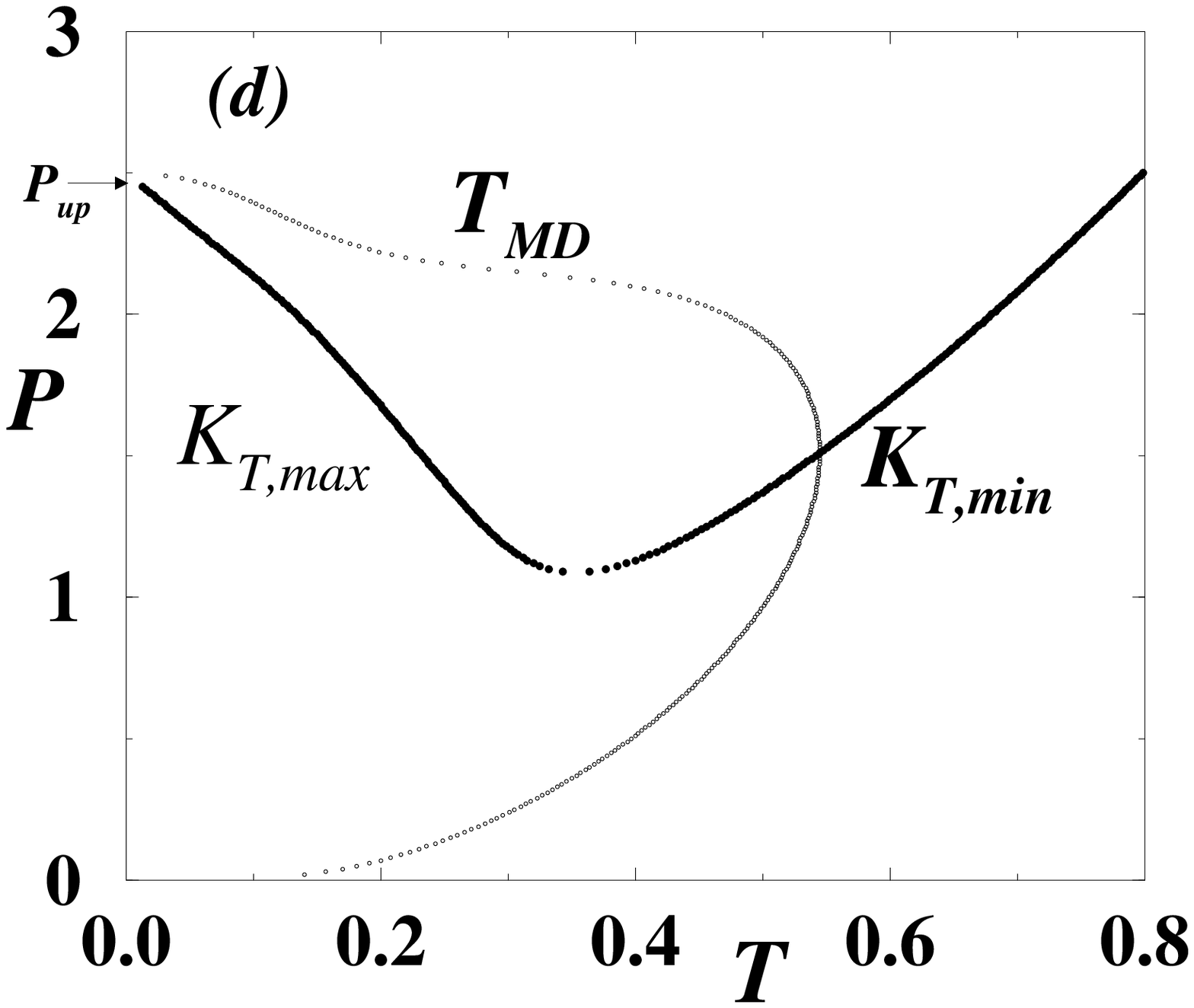}
	}
	  }

\caption{
(a) General form for the core-softened potential studied here. The
length parameters $a,b,c$ and energy parameters $\epsilon, \lambda$ are
shown. The dashed curve is the smooth version
\protect\cite{smoothformu}.  (b) Isobars (the average distance per
particle, $\ell$ versus $T$ at constant $P$) for the discrete $1d$
core-softened potential\protect\cite{smoothformu} with $P_{up}=2.5$ in
agreement with Eq.(\protect\ref{eqpu}). The $T_{MD}$ point is marked by
an open circle. (c) Isothermal compressibility for the discrete potential
along different isobars, with their maxima marked by filled circles. The
$K_T$ for $P_{up}$ diverges as $1/T$.  (d) The loci of $T_{MD}$ and
$K_T$ extrema for the discrete potential.}
\label{figpotlvst}
\end{figure}

The discrete form of the potential that we study is 
\begin{equation}
\label{uform}
 u(r)\equiv \left\{ \begin{array}{ll} 
	\infty &0<r<a \\ -\lambda \epsilon &a<r<b\\
	 -\epsilon &b<r<c\\
	0 &r>c \end{array} \right. 
\end{equation}
with $r$ being the inter-particle distance and $\lambda <1$ 
(Fig.~\ref{figpotlvst}(a))~\cite{smoothformu}. The model is exactly
solvable in $1d$, following the methods of~\cite{Takahashi,Yoshimura,Ben-Naim,Cho},
and the equation of state is
\end{multicols}
\noindent
$\underline{|\hspace{8.7cm}}$
\nopagebreak[4]
\begin{equation}
\label{1deos}
V(T,P)/N=a+\frac{1}{P}\left(k_BT+\frac{P\left[(b-a)(1-W^{\lambda})\Pi_b+\left((b-a)\Pi_b-(c-a)\Pi_c\right)(W-1)\right]}
	         {\Pi_a
	         W^{\lambda}+\Pi_b(1-W^{\lambda})+(\Pi_b-\Pi_c)(W-1)}\right).
\end{equation}
\nopagebreak[4]
\hspace*{0pt}\hfill$\overline{|\hspace{8.7cm}}$
\begin{multicols}{2}
\noindent
Here $V/N\equiv\ell$ is the average distance between nearest neighbors,
$\Pi_x\equiv e^{-\beta Px}$ ($x=a,b,c$), $W\equiv e^{\beta \epsilon}$
and $\beta\equiv k_BT$. The isobars (Fig.~\ref{figpotlvst}(b)) exhibit
two different types of behavior. For all $P$ larger or equal to an upper
boundary pressure $P_{up}$, $\ell=a$ at $T=0$, and $\ell$ increases
monotonically with $T$. For $P<P_{up}$, $\ell=b$ at $T=0$. The isobars
show a maximum and a minimum in $\ell$, which correspond respectively to
points of minimum and maximum density\cite{mindensnote}, bounding a
density anomaly ($\alpha_P<0$) region~\cite{Bell0,Bell}. There is a
discontinuity in $\ell$ at $P=P_{up}$ along the $T=0$ isotherm.

Next we study the isothermal compressibility
$K_T\equiv-V^{-1}\left({\partial V/\partial P}\right)_{T,N}$. We use
Eq.(\ref{1deos}) to calculate $K_T$ along isobars
(Fig.~\ref{figpotlvst}(c)).  The graphs show an anomalous region in
which $K_T$ decreases upon heating (for simple liquids $K_T$ increases
with $T$). We find the maximum value of $K_T$ grows as $P$ increases
towards $P_{up}$, and $K_T$ diverges as $1/T$ when we approach the point
$C'$ with coordinates $(T=0,P=P_{up})$ which we interpret as a critical
point~\cite{usvsstell}. Further, the locus of $K_T$ extrema joins the
point $C'$(Fig.~\ref{figpotlvst}(d)).  

We also study the $T_{MD}$ locus (Fig.~\ref{figpotlvst}(d)) and note
that the locus of $K_T$ extrema intersects the $T_{MD}$ locus at its
infinite slope point, a result that is thermodynamically
required\cite{Sastry}. Such a point on the $T_{MD}$ has been observed in
simulations which support the existence of a liquid-liquid phase
transition in supercooled water~\cite{criticalpoint}.

\begin{figure}[htb]
\narrowtext
\centerline{
\hbox  {
        \epsfxsize=4.0cm
        \epsfbox{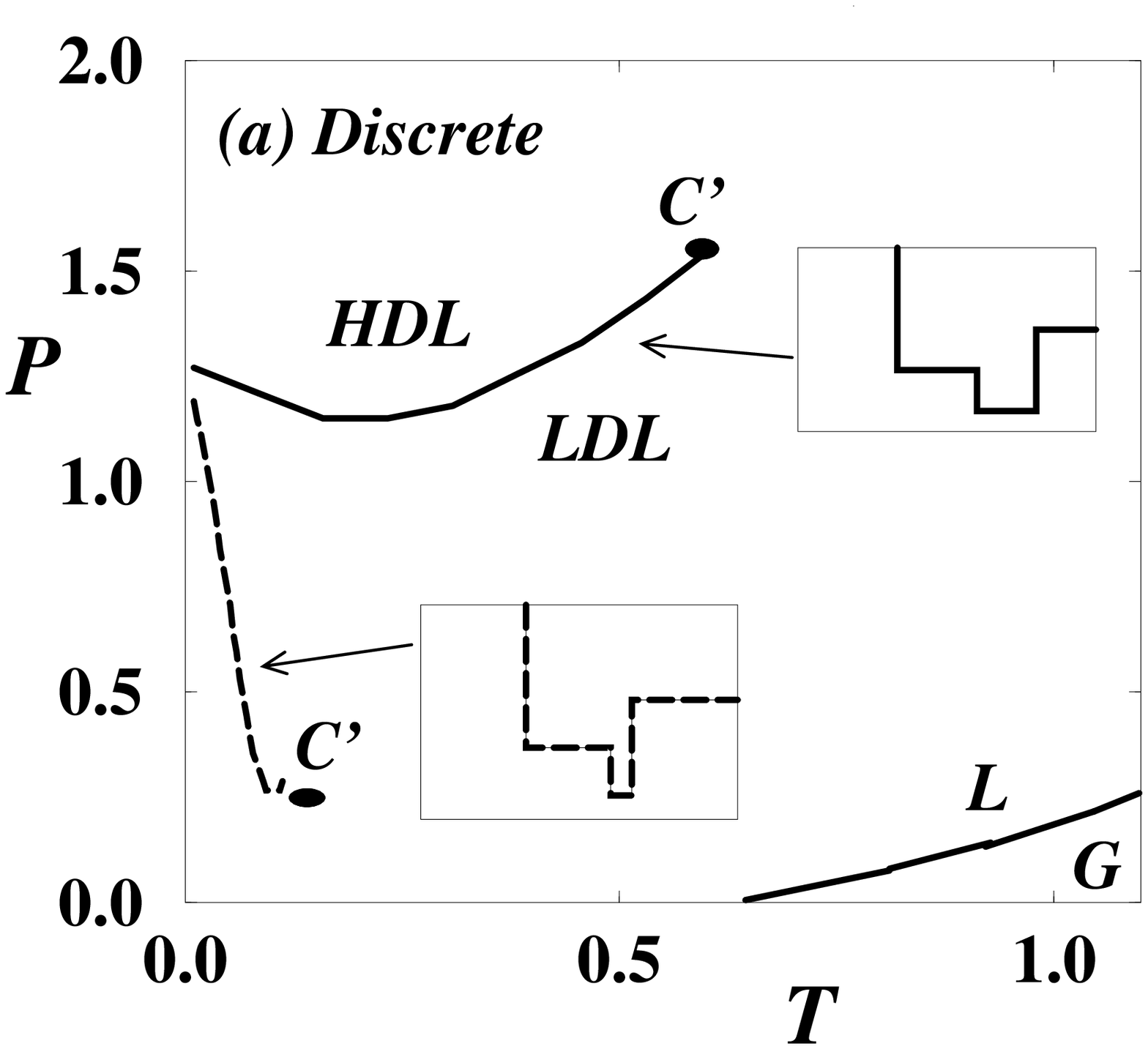}
        \hspace*{0.1cm}
        \epsfxsize=4.0cm
        \epsfbox{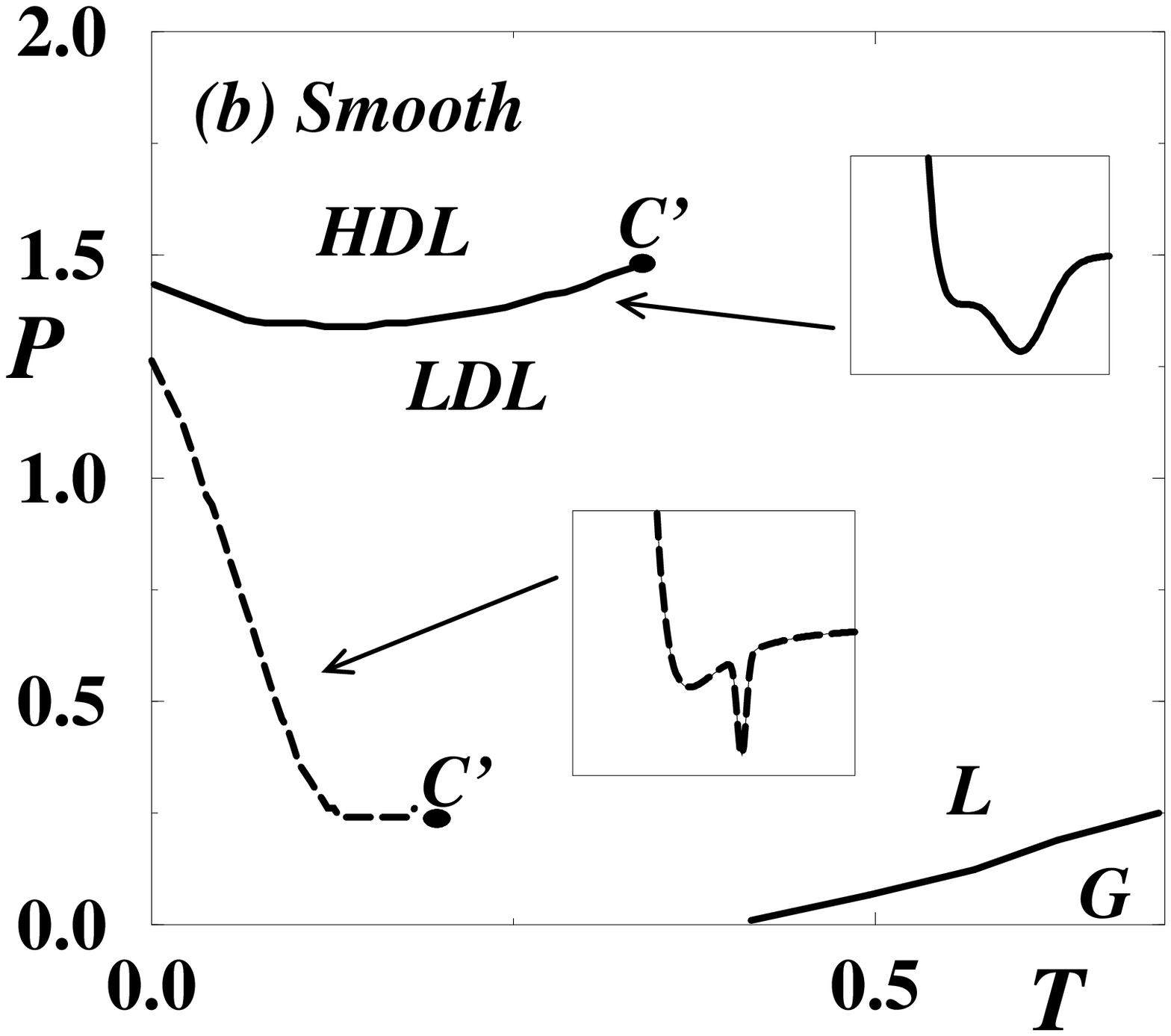}
       }
          }     

\caption{ 
(a) Solid lines correspond to the phase diagram from the cell theory method for the $2d$ discrete
potential with the same parameters as in 
Fig.~\protect\ref{figpotlvst} (solid
curve inset).  The lower solid line corresponds to the liquid-gas (L-G)
phase transition, with a  critical point outside the range of the
graph. The upper solid line corresponds to the HDL to LDL phase
transition, the slope of which changes from negative to positive. The
Clausius-Clapeyron equation relates the slope of a first order
transition line to the difference between the entropy and volume of the
phases by $dP/dT=\Delta S/\Delta V$. In the case of water, the open LDL
structure produced by highly directional hydrogen bonding has a lower
entropy. We can mimic this situation for our radially symmetric
potential by narrowing the well which greatly reduces the positively
sloped part of the transition line; the dashed line corresponds to the
HDL-LDL transition for a
discrete potential with $c=1.40001$.  
(b) Similar phase diagrams for the smooth version of the potential in
Fig.~\protect\ref{figpotlvst}(a)(solid lines).
The dashed line correspond to a potential with a much narrower well
($w=10,000$).}
\label{cellisos}
\end{figure}

We next consider the $2d$ case, for which an exact solution does not
exist. We use the spherical Lennard-Jones and Devonshire cell theory
method\cite{celltheory} which assumes that each particle is confined to
a circular cell, whose radius is determined by the average area per
particle, $v$. This method neglects the correlation between the
positions of different particles and assumes that the potential acting
on each particle is a result of interacting with all its nearest
neighbors smeared around its cell. The Helmholtz free energy per cell is
\begin{equation}
\label{eqfe}
h(v,T)=h_i(v,T)-k_BT\ln(v_f(v,T)/v),
\end{equation}
where $h_i(v,T)$ is the ideal gas free energy and $v_f(v,T)$ is the
free volume defined as
\begin{equation}
\label{freev}
v_f(v,T)\equiv \int_{\mbox{cell}}e^{-\beta u(x)}d\mbox{\bf x}
\end{equation}
with the core-softened potential used for $u(\mbox{\bf x})$ . For each
value $(P,T)$, we find the value of $v(P,T)$ by minimizing $h(v,T)-Pv$.
The resulting phase diagram (Fig.~\ref{cellisos}) has two lines of first order phase
transition, a low pressure line which is the liquid-gas phase transition
line terminating at a critical point $C$, and a high pressure line
that separates a low-density liquid (LDL) and a
high-density liquid (HDL) and terminates in a critical point $C'$.  This
picture holds both for the discrete and smooth versions of the
potential.  We note that the presence of $C'$ would imply an anomalous
increase in $K_T$ upon cooling when $C'$ is approached from higher
temperatures.

\begin{figure}[htb]
\narrowtext
\centerline{
\hbox  {
        \vspace*{0.1cm}
        \epsfxsize=7.0cm
        \epsfbox{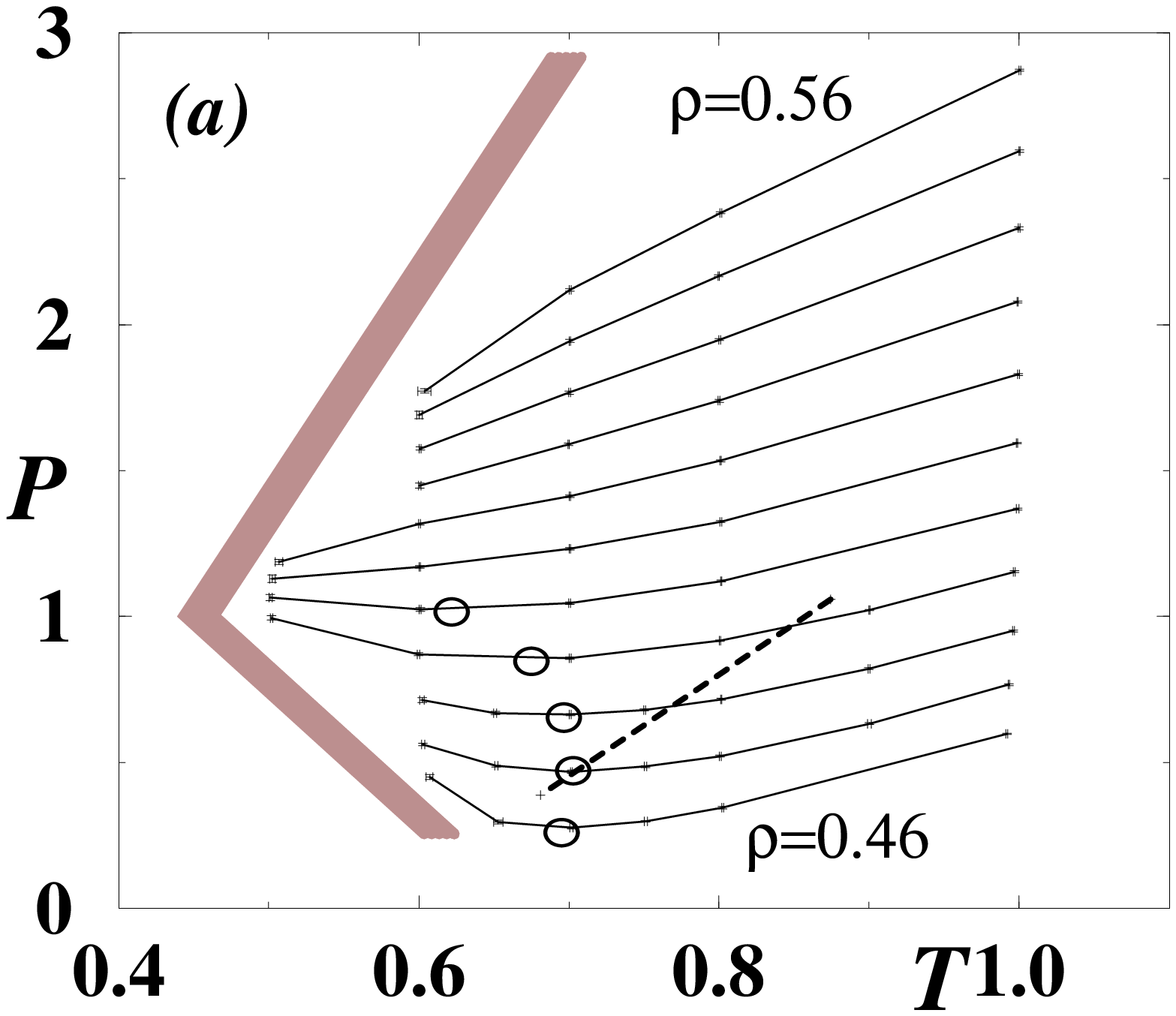}
       }
          }     
\centerline{
\hbox  {
        \epsfxsize=4cm
        \epsfbox{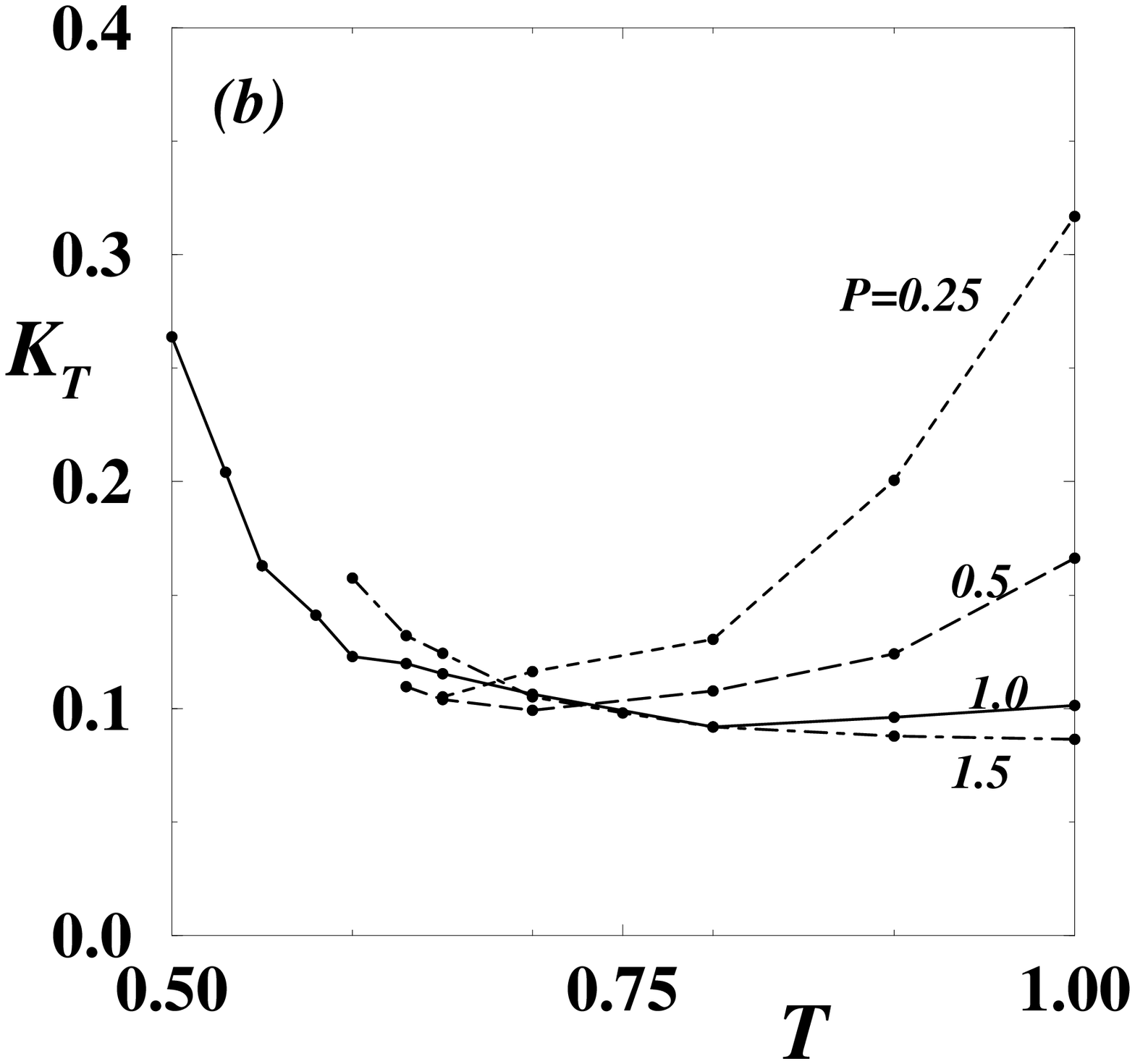}
        \vspace*{0.1cm}
        \epsfxsize=4cm
        \epsfbox{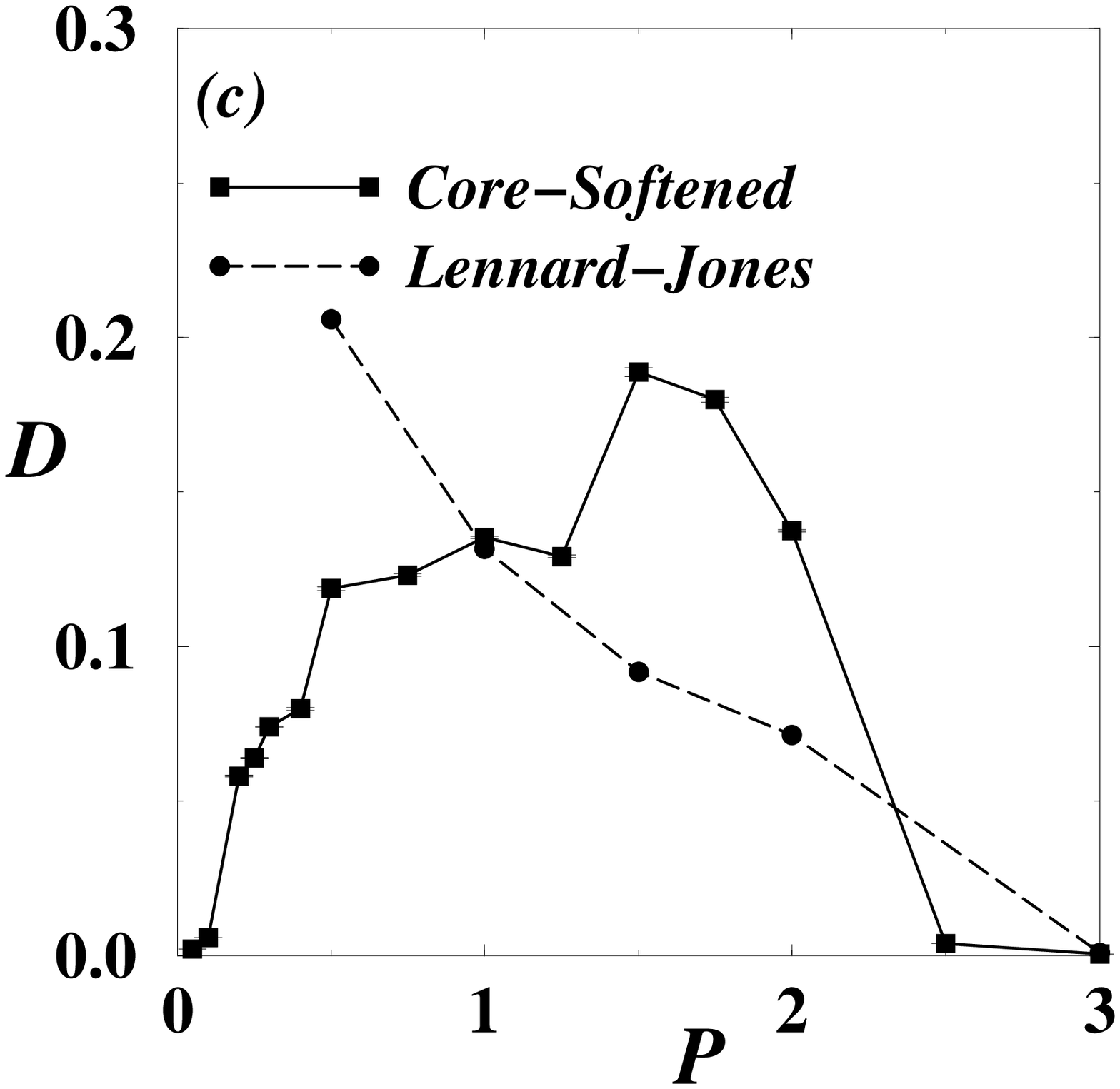}
       }
          }     
\caption{ 
MD results for the smooth potential with the same parameters as in
Fig.~\protect\ref{figpotlvst}\protect\cite{smoothformu}. (a) Constant
density curves with, from bottom to top, densities between $0.46$ to
$0.56$ in steps of $0.1$. The open circles mark $T_{MD}$ and the dashed
line is the locus of $K_T$ minima from part (b).  The thick gray line is
the approximate locus of the freezing points.  (b) Isothermal
compressibility along isobars. Except for the $P=0.25$ isobar, the
graphs show anomalous decrease upon heating.  (c) Diffusion
coefficient $D$ (slope of the mean square displacement as a function of
time) for different pressures at $T=0.65$, showing an anomalous increase
in the $P<1.5$ range. For comparison, we show $D/4$  for a
Lennard-Jones liquid at $T=0.7$, from simulation of $2304$
disks. The high pressure zero values in both graphs correspond to a
solid phase. }
\label{2disochores}
\end{figure}

In order to further investigate the system in the liquid phase we rely
on numerical molecular dynamics (MD) study of the discrete and smooth
versions of the potential (Fig.~\ref{figpotlvst}(a)).  We perform $2d$
MD simulations for a system composed of $N$ circular disks, in a
rectangular box of area $A$. To each disk we assign a radius equal to
half the hard core diameter $a$, and define the density $\rho$ as the
ratio of the total area of all the disks to the area of the box. For the
discrete version of the potential we use the collision table
technique~\cite{Allen-Tildesley} for $N=896$ disks and for the smooth
version of the potential, we use the velocity
Verlet integrator method~\cite{Allen-Tildesley} with $N=2500$. We set
the mass of the particles to be unity, while the units of length and
energy are scaled by $\epsilon$ and $a$ respectively. We choose the time
step $\delta t=0.01$~\cite{simulationspeed}. For each $T$, we first
slowly thermalize the system, using the Berendsen method of rescaling
the kinetic energy~\cite{Allen-Tildesley}, after which we perform the
simulation at constant $N,A$ and energy $E$. We fix $\rho$ by fixing $A$
and we start from $T=1$, lowering $T$ down to $0.4$ (in steps of $\Delta
T=0.1$ for larger $T$ and $\Delta T=0.05$ and $0.025$ in the vicinity of
freezing). As the initial configuration for each $(\rho,T)$, we choose
the equilibrated configuration of $(\rho,T+\Delta T)$. We simulate state
points along constant $\rho$ paths (isochores)
(Fig.~\ref{2disochores}(a)) and also along constant $P$ paths
(isobars)~\cite{isobarsims}. We use the isobar results to check the
values of $\rho(T,P)$ calculated from isochores, as well as to find
$K_T$ along isobars (Fig.~\ref{2disochores}(b))~\cite{ktdefp}.

The MD results are qualitatively equivalent for the discrete and smooth
versions of the potential. For the results of Fig.~\ref{2disochores} we
have used the smooth version. The minima in the $P$ versus $T$ isochores
(Fig.~\ref{2disochores}(a)) correspond to density maximum points
\cite{TMDformula}. We also find that along some of the isobars, $K_T$
increases upon cooling (Fig.~\ref{2disochores}(b))~\cite{Velasco}.  The
$T_{MD}$ locus possesses a point with infinite slope (as in $1d$), which
we verify by finding the intersection of the locus of $K_T$ minima with
the $T_{MD}$ line in Fig.~\ref{2disochores}(a)\cite{Sastry}. If we
assume that a metastable liquid critical point exists below freezing,
then by fitting the $K_T$ graph to a power law divergence, we can
estimate the critical point to be in the region $0.3<T<0.5$ and
$1.0<P<1.5$ which is in agreement with the cell-theory approximation.

We also study the effect of pressure on diffusion, and find that along
some isotherms, increasing pressure increases $D$, while for simple
liquids increasing pressure decreases $D$
(Fig.~\ref{2disochores}(c)). This anomaly occurs in the same region of
phase diagram where the density and isothermal compressibility anomaly
is observed.

\begin{figure}[htb]
\narrowtext
\centerline{
\hbox  {
        \epsfxsize=4cm
        \epsfbox{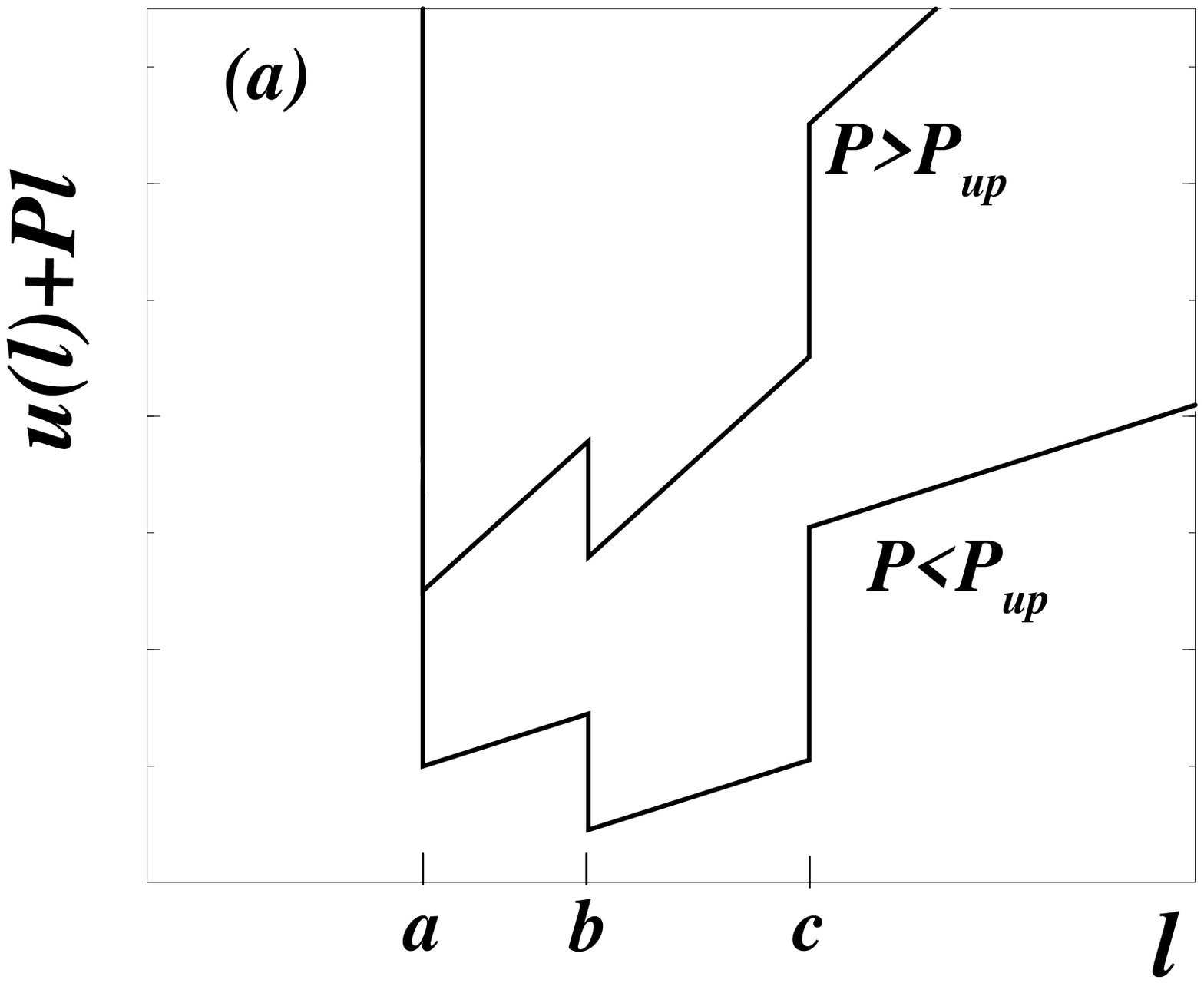}
        \hspace*{0.1cm}
        \epsfxsize=4cm
        \epsfbox{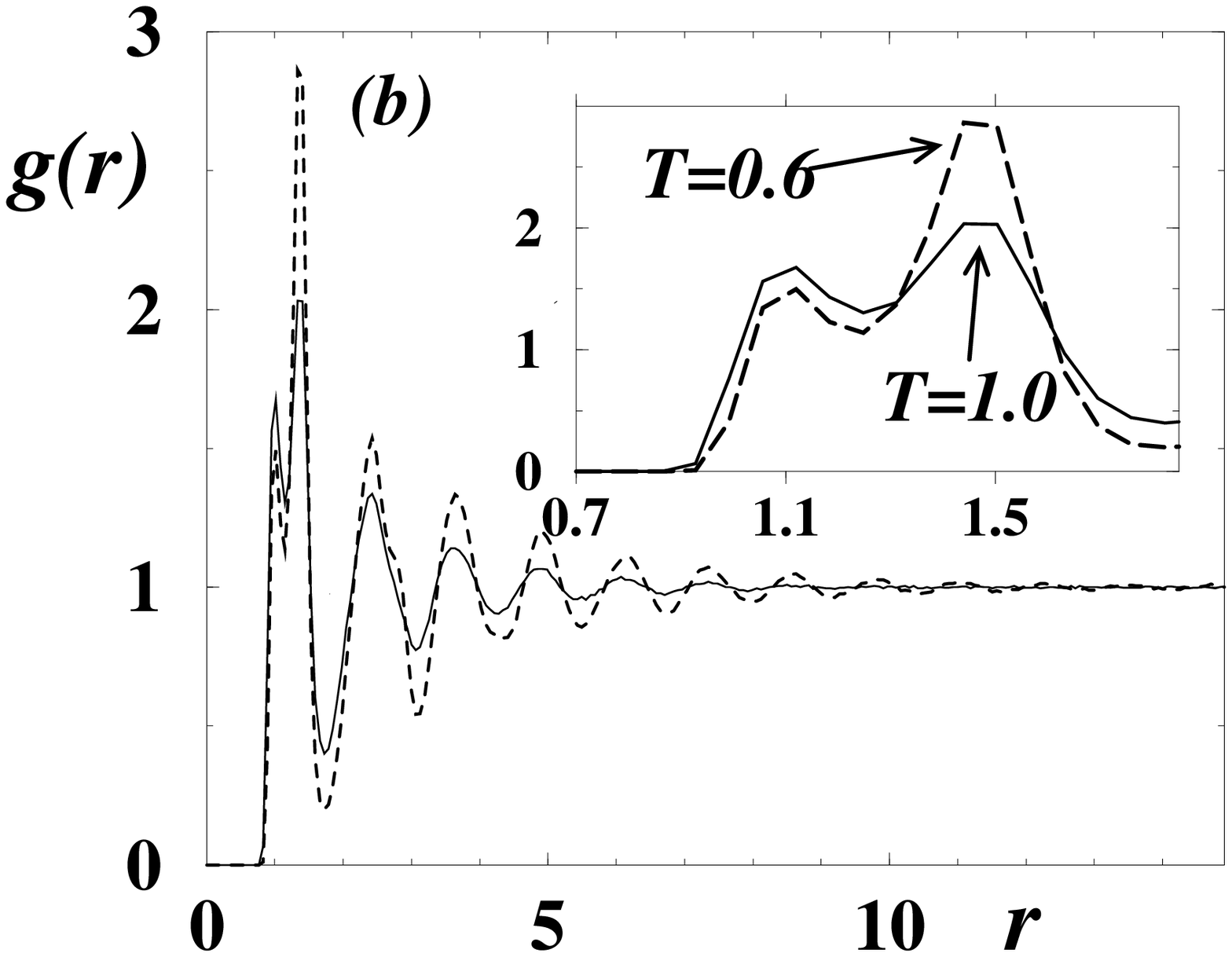}
       }
          }     
\caption{
(a) The 1d Gibbs free energy at $T=0$ as a function of the extra ``degree
of freedom'' $\ell$, for the discrete form of potential in
Fig.~\protect\ref{figpotlvst}(a).  The equilibrium value of $\ell(P)$ is
determined as the absolute minimum of this function, which is located at
$\ell=b$ below $P_{up}$ and at $\ell=a$ above $P_{up}$. (b) Pair distribution
function for $T=0.6$ and $T=1.0$ on the $\rho=0.48$ isochore of
Fig.~\protect\ref{2disochores}(a). The uniform $g(r)$ for large $r$ is a
sign of the liquid phase. The inset is a blow up of the first (split)
peak. The thick solid curve is for $T=1.0$ and the dashed curve is for
$T=0.6$, indicating that increasing $T$ lowers the total number of
particles in the open structure ($r\approx 1.5$) and increases the
number of particles in the dense structure ($r\approx 1.1$).
}
\label{structs}
\end{figure}

The anomalies can be related to the interplay between two local
structures, an open structure in which the nearest neighbor particles
are typically at a distance $b$ and a denser structure in which the
nearest neighbors penetrate into the softened core and are typically at a smaller
distance $a$. The configurations are determined by the minima of the
Gibbs free energy, $G(T,P)=U+PV-TS$ (where $G,U$ and $S$ are the Gibbs
free energy, internal energy, and entropy). Fig.~\ref{structs}(a)
shows the 1d free energy at $T=0$ for two different values of
$P$. The qualitative shape should not change for higher dimensions
and small $T$. For low pressures at small $T$, the open structure is
favored by the free energy. Increasing $T$ for these pressures will
increase local fluctuations in the form of dense structures which can
lead to an overall contraction of the system upon heating, causing a
density anomaly.  Increasing $P$ raises the relative free energy of the
open structure, until the dense structure will be the favored local
structure, as seen for $P>P_{up}$ in Fig.~\ref{structs}(a). At small $T$
this can lead to a first order pressure driven transition
(``core-collapse''\cite{Stell}), while for large $T$ the transition is
continuous. At $T=0$, the value of $P_{up}$ where the transition
occurs is where $G(T=0,P)$ is the same for the open and dense
structures, so
\begin{equation}
\label{eqpu}
P_{up}=-\frac{U_{open}-U_{dense}}{V_{open}-V_{dense}}.
\end{equation}
From Eq.(\ref{uform}) and Eq.(\ref{eqpu}) we find
$P_{up}={(1-\lambda)\epsilon/(b-a)}$ in 1d, which can also be derived
from Eq.(\ref{1deos})\cite{fnpu}.

To examine the transition from the open structure to dense structure in
$2d$, we study the pair distribution function $g(r)$ for the MD
configurations (Fig.~\ref{structs}(b)). The first peak in $g(r)$ splits
into two subpeaks, which correspond to the locations of the nearest
neighbors in the dense and open structures. As $T$ increases, the
open structure subpeak decreases while the dense structure subpeak
increases. We observe the same change with $P$ along the liquid
isotherms for small $T$. The uniform value of $g(r)$ for large $r$
confirms that all the state points of Fig.~\ref{2disochores}(a) are in
the liquid state.

We thank M.~Canpolat, S.~Havlin, B.~Kutnjak-Urbanc, M.~Meyer, S.~Sastry,
A.~Skibinsky, F.~Starr, G.~Stell and D.~Wolf for helpful discussions,
NSF for financial support.

\vspace*{1cm}

\end{multicols}

\begin{thebibliography}{99}
\bibitem{smoothformu} We also use a smooth form of the potential with
the form
$u(r)=4\lambda'\epsilon'(r^{-12}-r^{-6})-\epsilon'\exp[-(w(r-r_0))^n]$.
Unless stated otherwise, all 
of the results reported here are for a discrete potential with $a=1$,
$b=1.4$, $c=1.7$, $\epsilon=2$, and $\lambda=0.5$ and its smooth version
with $\epsilon'=1.7$, $\lambda'=1/1.7$, $w=5$, $r_0=1.5$, $n=2$
(Fig.~\ref{figpotlvst}(a)).
\bibitem{mindensnote}We focus attention on the maximum density point,
which is located at a higher temperature and has been observed in
several real materials, including liquid water, and in
simulations\cite{criticalpoint}. To our best knowledge, the minimum
density point has only been observed for the liquid tellurium alloy with
selenium (See ~\cite{Yoshimura} and references therein). In this $1d$
phase diagram, in addition to the the upper bound $P_{up}$, there is a
lower bound where the locus of density maxima ($T_{MD}$ line) terminates
by joining the locus of density minima. Below this lower bound the
isobars do not show a density anomaly.
\bibitem{Bell0} Similar behavior of isobars has been previously
reported for two-state lattice models \cite{Bell}. Also the existence of
the maximum density point for $P<P_{up}$ in $1d$  has been reported for
a double well potential (in which the potential has a barrier between
the two energy states\cite{Cho}).  The double well form of the
potential, because of its region of negative curvature, belongs to the
general category of core-softened potentials.
\bibitem{usvsstell} For the Stell-Hemmer potential with
a long-range attractive tail added, the line of first order transitions
extends up to a critical point. For our $1d$ short-range potential,
there cannot exist any critical point for $T>0$.
\bibitem{fnpu}In higher dimensions, Eq.(\ref{eqpu}) helps 
estimating the pressure region in which we expect to observe the
anomalies.
\bibitem{TMDformula} Since $\left({\partial P}/{\partial T}\right)_V={
\alpha_P}/{K_T}$, a minimum along the isochore implies $\alpha_P=0$.
\bibitem{ktdefp} We calculate $K_T$ using $K_T=\lim_{q\to0}S(q)/\rho k_BT$,
where the structure function $S(q)$ is the Fourier transform of the pair
distribution function $g(r)$.  We also use the relation $K_T=V<(\delta
n)^2>/(k_BT<n>^2)$, where $<n>$ and $<(\delta n)^2>$ are average number
density and its variance averaged over boxes of edges equal to $1/15$
of system edges. The results of the two methods agree.
\end{thebibliography}
\end{document}